# Vector-borne threats: Sustainable approaches to their diagnosis and treatment

Areesha Naveed[† 1], Ayesha Haidar[†2], Rameen Atique[3], Arshi Saeed[4], Bushra Anwar[5], Ambreen Talib[6], Uzma Bilal[7], Javeria Sharif[8], Ayesha Nadeem[8], Sania Tariq[9], Ayesha Muazzam[10*], Abdul Samad[11*]

[1,2,3,4,5,6,8] Department of Pathobiology and Biomedical Sciences, Muhammad Nawaz Shareef University of Agriculture, Multan, 25000, Pakistan.

[7] Department of Zoology, The Women University Multan, 60000, Pakistan

[10] Department of Animal and Dairy Sciences, Muhammad Nawaz Shareef University of Agriculture, Multan, 25000, Pakistan.

[9, 11] Division of Applied Life Science (BK21 Four), Gyeongsang National University, Jinju 52852, Korea

**Abstract:**

Arbovirus is a vital, life-threatening disease worldwide and continues to be a significant problem while the world is dealing with the major coronavirus (COVID-19) pandemic. Vectors, mostly mosquitoes and ticks, transmit this disease. Dengue fever, chikungunya, and Zika viruses are the major threats because of their high incidence, public health burden, and clinically significant disease spectrum. These vector-borne disease causes one-fourth of annual deaths, leading to various infectious diseases. The arbovirus represents eight different families and 14 genera; most viruses belong to the family Bunyaviridae, and some also belong to Togaviridae, Reoviridae, and Flaviviridae. The arbovirus disease was isolated first in tropical and subtropical regions of South America and Africa and has high significance because of suitable environmental conditions for virus transmission and vector expansion. Its transmission cycle ranges from simple to highly complex. DENV is the most prevalent, results in febrile illness, and has transmission in 128 different countries. CHIKV causes infection in asymptomatic people, and the problems include nephritis, arthritis, myelitis, and acute encephalopathy. ZIKV-infected 80% of people are asymptomatic and may cause rashes, myalgia, fever, headache, and conjunctivitis. Vaccines for DENV are not clinically available; it is a primary arboviral infection in the world nowadays. The exposure of arbovirus diseases continues to be a global health problem regardless of continuing efforts. This review article will overview major arbovirus diseases and their diagnosis, treatment, and prevention strategies.

**Keywords:**

1. Introduction:

Arboviral diseases are infections caused by viruses that develop in people by biting infected arthropods (insects) such as mosquitos, ticks, and aphids. Insect and arthropod vectors have been responsible for the presence of pathogens in humans, animals, and insects that cause high rates of morbidity and mortality throughout history. The epidemic of arboviral diseases affects health and public affairs globally, mainly in underdeveloped nations [1].In tropical areas, arboviruses are very deadly diseases. It involves various procedures like the enzootic cycle, which includes domestic animals (mice, monkeys, mosquitoes); the urban epidemic cycle transmitted by Aedes aegypti mosquito to humans also, which results in human amplification; and the rural epizootic cycle transmitted by pigs, horse to mosquito and then to human [2]. There are 534 viruses in the International Catalog of Arbovirus, which have vertebrate animals as their primary host, and 134 viruses result in human infection. There are also eight families and 14 genera of Arbovirus infection [3]. Some diseases cause mild illness, while some are fatal and cause significant mortality. Various vector-borne diseases, including malaria, dengue, yellow fever, plague, and leishmaniasis, altogether caused death in humans in the 17$^{th}$ century. However, in the 20$^{th}$ century, various strategies to reduce the mosquito population have helped reduce the impact of these diseases on human health. Mosquitoes and other arthropods, including viruses, spread many significant infections to medicine.

A diverse range of medically essential arboviruses comprises the family Flaviviridae, which includes West Nile virus (WNV), Dengue virus (DENV), Japanese encephalitis virus (JEV), Yellow Fever virus (YFV) and Zika virus (ZIKV), Tick-borne encephalitis virus (TBEV), Togaviridae family include Chikungunya virus (CHIKV), Western equine encephalitis virus (WEEV), Eastern equine encephalitis virus (EEEV), O'Nyong-Nyong virus (ONNV), Sindbis virus (SINV) and Bunyaviridae family include Rift Valley fever virus (RVF), Sandfly fever virus (SFSV) are among these arthropod-borne viruses, also known as arboviruses [4]. The classification of arbovirus is shown in **Fig 1**. Different zoonotic viruses like Vesicular stomatitis virus (VSV), African horse sickness virus (AHSV), Epizootic hemorrhagic disease virus (EHDV), and blue tongue virus (BTV) do not cause disease in humans but cause serious diseases in animal populations and leads to economic disruptions. Various arbovirus causes clinical symptoms like hemorrhagic fever, headache, pain, febrile illness, myalgia, and, in severe cases, death [3]. Various factors like human behavior, climate, and environment affect the transmission of arboviruses. The approaches to lessen the thrust of arbovirus diseases include controlling vector transmission and the Aedes mosquito population [5].

**Fig. 1. Classification of Arboviruses**

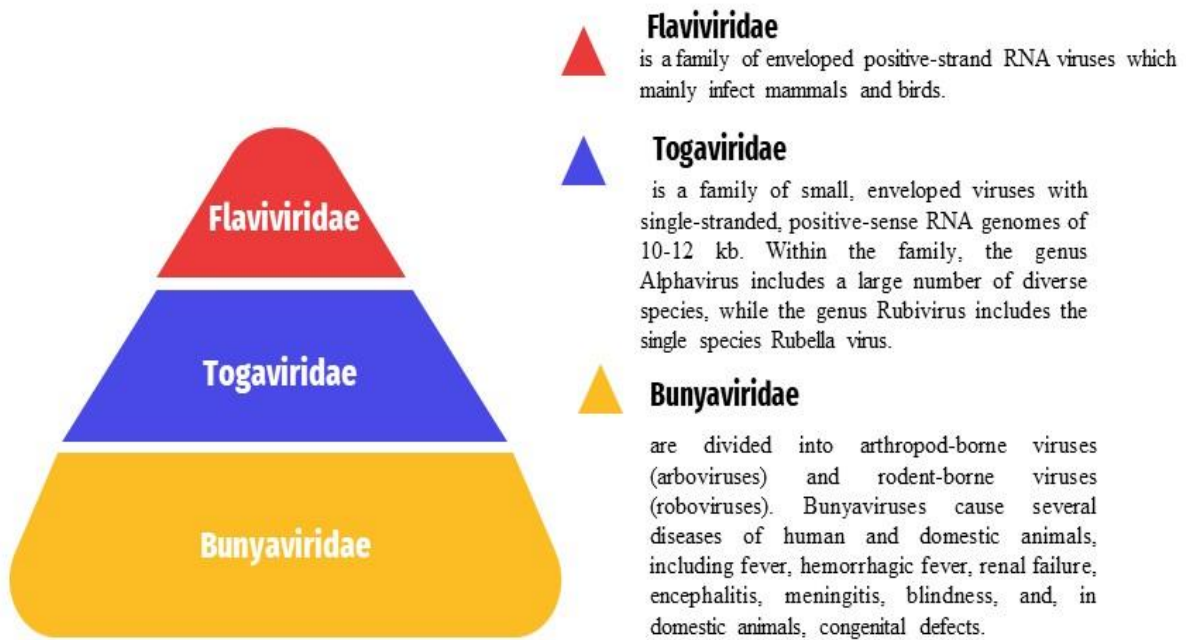

**Flaviviridae**
is a family of enveloped positive-strand RNA viruses which mainly infect mammals and birds.

**Togaviridae**
is a family of small, enveloped viruses with single-stranded, positive-sense RNA genomes of 10-12 kb. Within the family, the genus Alphavirus includes a large number of diverse species, while the genus Rubivirus includes the single species Rubella virus.

**Bunyaviridae**
are divided into arthropod-borne viruses (arboviruses) and rodent-borne viruses (roboviruses). Bunyaviruses cause several diseases of human and domestic animals, including fever, hemorrhagic fever, renal failure, encephalitis, meningitis, blindness, and, in domestic animals, congenital defects.

2. **EvolutIon of arboviruses:**

Various arbovirus outbreaks in different countries mainly result from climate changes and landscapes. All the arboviruses reservoir hosts and infection vectors are discussed in detail in **Table 1**. In the Pre-20$^{th}$ century, there was an epidemic of diseases like the Yellow Fever virus in Africa, but it was not known as arbovirus disease because no such term originated. In the 20$^{th}$ century, Finlay and Walter originated the transmission of the Aedes mosquito that resulted in YFV. This virus was not eliminated by vector control and spread to Caribbean Island and Mediterranean Europe, resulting in the fatal outbreak of YFV [6].

In 1978, the first indication that mosquitoes may carry human infection was discovered filarial worm transmission of mosquito by Sir Patrick Manson. In 1881, Carlos speculated that the agent responsible for yellow fever was a mosquito. In 1900, James Carroll, serving in Cuba, let Jesse Lazear lay a contagious Aedes aegypti mosquito on Carroll's arm, allowing the insect to feed on human blood. Carroll was thought to have contracted yellow fever, but he recovered. The research conducted by Walter, Carroll, and Lazear proved that mosquitos were the means of transmitting the yellow fever virus (YFV). Lazear passed away from yellow fever on September 25, 1990, before the last investigation was finished [7]. In 1950, the word "Arbovirus" was presented for the disease transmitted by arthropod vectors such as ticks, sandflies, and mosquitoes. After a 35-year hiatus, the alphavirus O'nyong- nyong virus (ONNV) revived in Uganda in 1996 and caused illness. The best case of arbovirus documented due to climatic change is WNV. 1999, it was introduced in different countries like New York, Canada, and Central and South America, increasing its geographic range. WNY affected more than 40,000 people in the U.S. and over 250 species of birds. In 2000, cases of RVFV were disclosed for the first time far from Africa. A related virus to ONNV and RVFV first appeared in Kenya in 2004 and proceeded to Indian Ocean Island in 2005. Recent initial appearances in Europe of other arboviruses, such as BTV and the Usutu virus, have substantially impacted cattle and wildlife populations [8].

In recent decades, many arboviruses have come to light and are recognized as major public healthcare threats in the 21$^{st}$ century. Dengue is a major arthropod-borne illness; more than 380 million dengue cases are thought to occur in humans annually. It has been challenging to assist the use of vector control techniques as the control strategies are the only means to prevent arbovirus [9]. The latest cases of arboviruses demonstrate the need for greater awareness of the parameters that influence the arbovirus emergence and evolution and the damage caused by viruses like DENV, YFV, and WNV, which are widespread worldwide. Various ongoing research concentrates on understanding the epidemiology, pathogenesis, and factors influencing arbovirus evolution and developing antiviral drugs and vaccines for arboviral diseases [8]. COVID-19 has various impacts on arboviral diseases. At first, travel limitations brought on by COVID-19 quarantines may be seen as a good indication, but they will be overwhelmed by some adverse effects of the virus. COVID-19 and dengue have caused problems like making it harder for doctors to diagnose the illness properly, as dengue is the common cause of illness among travelers, and producing false-negative results in patients who already have DENV antibodies, which results in interaction between DENV and SARSCoV-2 [10].

The maximum likelihood method (MCL) approach discovered each virus's evolutionary history. The trees were constructed using BioNJ algorithms and neighbor-joining on bilateral distance computed using the MCL approach, and then the hierarchy with the highest log probability value was selected [11]. According to phylogenetic studies of Arbovirus, purifying selection is a positive factor in arbovirus evolution that examines the fraction of changes throughout time. The widely recognized explanation states that arbovirus loses its potential as host specialists because cycling between hosts favors specialists [8].

**Table 1: Various arboviruses with infection vectors and reservoir hosts**

| Arboviruses | Infection vectors | Reservoir hosts | References |
|---|---|---|---|
| Yellow fever virus | Aedes specie mosquitos | Human beings, Monkey | [12] |
| West Nile virus | Culex specie mosquitos | birds | [13] |
| Dengue virus | Aedes aegypti and aedes albopictus mosquitos | Human beings, monkey | [14] |
| O'nyong-nyong virus | Anopheles gambiae mosquitos | Unknown | [15] |
| Chikungunya virus | Aedes aegypti mosquitos | Bats, human beings, wild primates | [16] |
| Japanese encephalitis virus | Culex quienquefasciatus mosquitos | Wading birds, pigs | [17] |
| Rift Valley fever virus | Culex and aedes specie mosquitos | Buffaloes, rodents, sheep, cattle | [18] |
| Tick-borne encephalitis virus | Hard ticks | Yellow-neck mice, voles, insects | [19] |
| Zika virus | Aedes specie mosquitos | Sheep, goat, cow, rhesus monkey, rodents | [20] |
| Sandfly fever virus | Sandflies | Human beings | [21] |
| Vesicular stomatitis virus | Sandflies, midges, and Aedes mosquitos | Cattles, horses | [22] |
| St. Louis encephalitis virus | Culex tarsalis mosquitoes | Birds such as sparrow, robin | [23] |
| La crosse encephalitis virus | Aedes triseriatus mosquitoes | Squirrels, chipmunks | [24] |

## 3. Major arbovirus diseases:
### 3.1 Flaviviridae:

The Flaviviridae family was early classified as "Group B Arboviruses" with both Flavivirus and Alphavirus genus, but later, Alphavirus belonged to the Togaviridae family. In 1980, Porterfield presented a progressive association between arthropod and non-arthropod flaviviruses [25]. The Flavivirus genome is a single-stranded, positive sense RNA enveloped virus with 68 species and host-specific (mammals, birds). It has a single open reading frame bound by 5'-and 3'-end antisense region. These viruses have a high impact on public health and carry more than 70 species. The replication of flavivirus, which has structural and non-structural proteins, occurs by combining antigenomes as a model for RNA display. The enzymes used for replication are RNA helicase, serine protease, and RNA dependent- RNA polymerase [26].

Culex mosquitos transmit West Nile Virus, the first cause of infection in North America in 1999 and 1937, first recognized from contaminated persons in Uganda. It involves bird-to-human transmission and infects the host. Its incubation period ranges from 3 to 14 days [27, 28]. Dengue Virus has four serotypes (DENV 1-4) transmitted by Aedes mosquito and involves mild-to-severe illness. When a person is infected with DENV serotypes, there is an increased risk of developing dengue disease [29]. A previous study of DENV-2 showed complete genome sequences after alternative passage in mosquito and mammalian cell lines, and DENV-2 shows fewer genetic changes in accord sequence from mosquito cell-derived virus. The assumption that cycling alone leads to constraints on accord changes is not supported by the result of in vitro flavivirus research, nor the occurrence of significant fitness trade-offs from host cycling supported. The study shows that host transitions, at least in the case of VSV, do not permanently restrict genetic change, nor does cycling constrain host-specific adaptations through cycling [8].

Japanese Encephalitis Virus is a single-stranded envelope with mosquitoes as vectors and pigs as reservoir hosts. The first case of JEV was reported in 1871 in Japan. It has four genotypes with unknown origins. In 1952, a stain of JEV was confined from a patient that appeared as the fifth genotype due to restricted evolutionary verification [30, 31]. Yellow Fever Virus is a life-threatening disease from the 18th century when there was no preventive measure to control its spread in Africa and America. YFV, also known as hemorrhagic fever, is transmitted by mosquitoes and monkeys and uses humans as hosts. There is no treatment for YFV, so vaccines like YF-17D are used as preventive measures. [12, 32]. Zika Virus was first isolated in 1947 from rhesus monkeys and infected people of Asia, Africa, and America with Aedes aegypti and Aedes albopictus, as shown in **Table 1**. In 1954, the first human case of ZIKV was recognized from a Nigerian female, and a further two cases of human ZIKV infection were identified in Nigeria that were verified by a boost in serum-neutralizing antibodies [33].

### 3.2 Togaviridae:

The Togaviridae family consists of 4 genera: Rubivirus, Alphavirus, Arterivirus, and Pestivirus. All virus genomes are 9.7 to 11.7 kb in size, round-shaped, 50-80 nm in diameter, and two or more polypeptides surrounded with exterior prognosis. Togaviridae is a positive sense with single-stranded RNA, enveloped, and its RNA is polyadenylated at the 3' end and has capping on the 5' end with non-structural protein in the form of polyprotein [34, 35]. Unlike the Flaviviridae or Picornaviridae, in which proteolytic processing is used to generate protein, the Togaviridae generate the viral protein by mRNA transcription in an assessable and temporary way [35]. The alphavirus structure has E1 and E2 glycoproteins in its lipid bilayer, and E1 and E2 glycoproteins are matched together and visible as spikes in a well-organized arrangement.

Alphavirus has 27 species with the same antigenic area on the capsid, but further diagnostic tests have evolved them. Antibodies and interferon are used for improvement and recovery and are bound to paired Alphavirus. Rubivirus has only one species; the Rubella virus and Pestivirus contain only animal infectious agents. The most critical Alphavirus disease is CHIKV, a positive sense RNA virus with polypeptides having four non-structural proteins and five structural proteins. The antigenic complexes have four important species: Western equine encephalitis, Eastern equine encephalitis, Semliki Forest virus, and Venezuelan equine encephalitis [36].

### 3.3 Bunyaviridae:

Bunyaviridae is a family of 300 viruses transmitted mainly by arthropods and includes viruses of four different genera: Bunyavirus, Phlebovirus, Nairovirus, Hantavirus, and Uukuvirus. Bunyavirus is a single-stranded RNA genome with a negative sense, characteristic rounded with a diameter of about 80-120 nm. The RNA genome consists of three fragments: large (L), Medium (M), and small (S), and these fragments have proteins necessary for reproduction and are organized in point- a hydrogen-rounded structure [37]. The viruses included in Bunyavirus are Rift Valley fever (RVF), La crosse virus (LACV), and Crimean-Congo hemorrhagic fever (CCHFV). Hemorrhagic fever associated with Bunyavirus results in a severe death rate in Asia [38].

RVF virus is a disease in humans and animals transmitted by mosquitoes insulated by more than 30 species. It causes infection by direct contact with infected animals. In 1951, a sudden disease outbreak occurred in South Africa, resulting in the death of thousands of animals [39]. Tree-hole mosquitoes mainly transmit LACV and are the most infective part of the California encephalitis serotype. According to the CDC, there are 25- 150 cases of LACV in humans, primarily children, in the USA each year. Very acute manifestation can result in long-term neural damage and, in a few cases, cause death [40]. CCHFV is a disease transmitted by ticks from the Nairovirus genus breakout in 1940. CCHFV has been isolated from sheep, mice, and dogs. There is only slight information about the disease development due to the lack of specifically designed laboratories and animal representatives [41].

### 4. Epidemiology:

The epidemiology of arboviral diseases is mainly based on the particular virus and its vector. Arboviruses are mainly prevalent in tropical and sub-tropical regions like African and Asian countries. The geographic location where arboviruses are prevalent significantly affects their epidemiology [42]. Arbovirus disease epidemic appears in various unaffected areas of the globe due to various factors like changes in the genetic structure of the virus, changes in vectors/host populations, and changes in weather conditions (temperature, rainfall, and humidity). Zika and West Nile viruses are mainly spread due to climate changes in moderate environments [43]. Environmental variation and human habits have a crucial effect on the disposition and transmission of arbovirus diseases. These diseases can be identified by improvements in various factors like records of animal bites, travel records, understanding of monitoring data, and current epidemic trends in society [44]. Yellow fever is a zoonotic disease transmitted by the bite of the Aedes species mosquito that mainly affects people in tropical and sub-tropical areas. The endemic region of yellow fever includes 13 countries in Central and South America, and the mortality rate is around 30,000 to 60,000 annually. Every vulnerable country should have one laboratory to perform blood tests for yellow fever [45]. Several outbreaks of YFV have been reported from the 15th century to the late 1990s; nowadays, 29 African countries have been infected by YFV [1]. The WHO explains that 95% of people were vaccinated against yellow fever in Ghana, but the outbreak of yellow fever occurred between 2021 and 2022. All the past cases of yellow fever in Ghana from 1910 to 2022 were compared with the help of multiple sources like WHO reports and literature, and it shows significant changes in epidemiology and geographic distribution of YF outbreaks over the past few centuries [46].

Dengue virus is the most severe vector-borne disease, with 390 million cases reported yearly, and the primary threat regions in America and Asia [42]. Dengue has been spreading to new regions since 2000 and is a significant threat in urban and rural areas where the Aedes mosquito is common [47]. Dengue cases also increased 8-fold to over 4.2 million in 2019 [10]. From 2010 to 2021, various travel-related dengue cases were reported in the United States, but later, the number of cases was reduced due to the COVID-19 epidemic [48]. Chikungunya virus disease was not recognized before, but its epidemiology has grown remarkably. Outbreaks of CHIKV have been reported in both tropical and temperate regions of Asia, Africa, Caribbean and America [1]. A cross-sectional study was conducted on chikungunya virus in Northern Peru from 2020 to 2021. In 2015, the first case was reported in Peru. Serum samples were then collected from patients, and real-time RT-PCR with a probe was performed to detect CHIKV. A total of 1047 patients with AFI volunteered for CHIKV detection. Of 1047, 130 had chikungunya identification, and 84 had RT-PCR diagnosis. The clinical symptoms of CHIKV disease might overlap with other AFI causes, so it is necessary to employ diagnosis tools like serological and molecular tests, primarily in areas where other arboviral diseases are predominant [49].

Like DENV, YFV, and CHIKV, the Zika mosquito also transmits the Aedes virus. On the contrary, the Zika virus is also transmitted from mother to infant during pregnancy and sexual contact. Zika virus was first identified in Uganda's Zika forest in 1947, and since then, it has been reported in many African and Asian countries. In 2015, the Zika virus was exposed in America, leading to an international health crisis [1]. The discipline of epidemiology accuracy has evolved into various challenges, such as creating precise diagnostic techniques for the infected person and more effective public-level treatment and monitoring programs [50]. The West Nile virus is a significant cause of infection of arthropod-borne virus (arbovirus) in the United States. CDC reported all the analyses and cases of WNV were reported from 432 countries. The 95% of neuroinvasive cases were described as neurologic illness, meningitis, encephalitis, and acute paralysis. Jamestown Canyon virus disease was reported in New Hampshire, affecting 0.29 per 100,000. Various factors like human behavior, weather, and vector abundance can affect where epidemics appear [51].

## 5. Vector mode of transmission:

The most common transmission vector of arboviral diseases like ZIKV, DENV, and CHIKV is the mosquito of the Aedes species. Aedes. aegypti is a tropical mosquito that feeds mainly on humans, takes several bites in a single meal, and does not survive in cold weather. Aedes. Albopictus is a temperate mosquito that is very aggressive and survives in both rural and urban areas, as shown in **Fig 2.** The culex species mosquito feeds mostly on birds instead of humans and sometimes drinks human blood, resulting in deadly infection [52]. The modes of transmission usually used for arbovirus diseases are horizontal and vertical [53, 54]. Horizontal transmission is the process in which the viruses are transmitted to individuals of similar breeding, while vertical transmission is where the virus spreads from the mother to their progeny. These two transmission modes altogether help the viruses to survive in any circumstances. Arbovirus epidemiological cycles may comprise complex transmission pathways, including animal host and blood-feeding arthropods, generally known as vectors, including mosquitoes and ticks [55].

**Fig. 2. Vector transmission cycles**

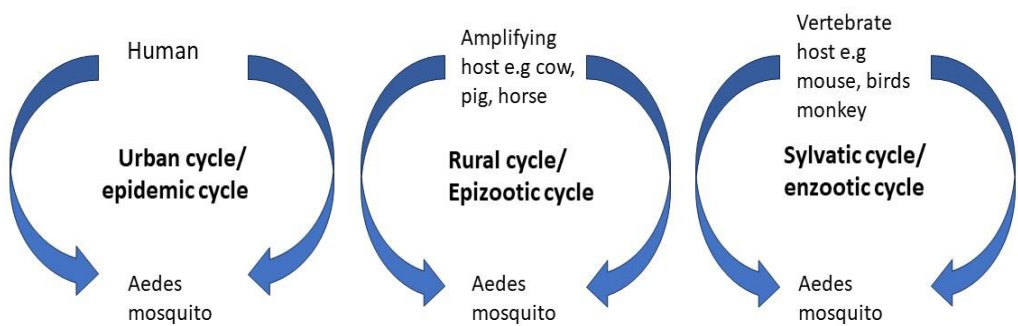

Arbovirus differs from other animal viruses as it is transmitted to animals by hematophagous vectors, biological transmissions involving three essential elements (vector, virus, vertebrate). The direct transmission method is commonly used in all arbovirus groups, and the transmission passage is oral, nasal, carnal, and skin disclosure with cuts, scraps, and tendon tissue. Vaporizers can also contaminate vertebrates with dengue virus (DENV). The oral transmission of arboviruses like YFV, JEV, and RVFV to mosquito larvae can happen in dirty water. Mosquitoes such as Aedes and anopheles mostly prey on mammals, while culex species prey on birds and small mammals [56].

Zika virus causes asymptomatic infection but, in some cases, results in neurological illness in mature persons due to congenital disabilities in infants and has been regarded as a public health threat [57]. The transmission cycle of the Zika virus is sustained by the sylvatic cycle and urban cycle. The sylvatic cycle involves transmission between mosquitoes and primates, and the urban cycle involves transmission between dogs and other animals/humans [58, 59]. CHIKV results in postpartum vertical transmission from the infected mother to the child, causes child death, brain problems in infected newborns, and genetic diseases. CHIKV is transmitted mainly by Aedes mosquitoes, but rodents and cattle are also associated with virus transmission. The widespread hosts of CHIKV are monkeys and primates [60]. YFV involves various Aedes species mosquitoes for transmission cycle in African countries [61]. YFV is also preserved in enzootic cycles that include monkeys and mostly have no symptoms [57]. Arbovirus diseases like DENV, YFV, CHIKV, And ZIKV are transmitted by Aedes mosquitoes, as shown in Fig 2.

Mosquitoes may also develop diverse defense mechanisms to detect viruses. All insects develop an innate immune pathway against contagious agents, the RNAi pathway [5]. Innate immunity is developed by perceiving pathogen-associated molecular patterns (PAMP) by host-pattern recognition receptor (PRR). The fruit fly Drosophila melanogaster is a template insect for studying anti-arboviral protection in mosquitoes. Gene expression is silenced by adding double-stranded RNA with a linked sequence, presented in nematode Drosophila and Caenorhabditis in 1998. The anti-viral RNAi response to mosquitoes provides a method to control arboviral transmission and replication by detecting double-stranded RNA [62].

The disease spreads, and virus infection to the host can be caused by the salivary gland of a mosquito. Mosquito saliva has various functions, such as pain repression, reduced inflammation, and anti-coagulation. The host immune response is also modified by disease transmission with saliva [63]. Arbovirus infects mosquitoes when they feed on infected hosts. The virus enters the mosquito midgut with a blood meal. There is selective selection against genotypes that are poorly suited to midgut activity. If so, it is unclear what trait the chosen population possesses that allows it to be well suited to the environment of the mosquito midgut [52].

6. **Signs and symptoms:**

More than 50% of viruses that cause human diseases belong to the flavivirus family, including the dengue virus and Zika virus. Some viruses also belong to the Togaviridae family, like the Chikungunya virus. In the early phase, it is tough to differentiate between these diseases. Thus, a clinical diagnosis is necessary that is based on signs and symptoms [54]. All the viruses have similar symptoms, resulting in high fevers, which creates a challenge in identifying the virus [64].

The incubation period of the chikungunya virus ranges from 2-7 days, after which viremia occurs (the virus appears in the blood), and symptoms progress. Its symptoms include high fever, arthralgia, pain, skin and maculopapular rashes, and ulcers. Joint pain, mainly in the fingers, ankles, and wrists, caused by the chikungunya virus can be very acute. The patients do not develop severe infections but sometimes show respiratory or vascular signs [65]. If there are no restrictions, acetaminophen should be used to treat persistent joint discomfort caused by a chikungunya infection [66].

The incubation period of dengue virus ranges from 3-14 days and has a febrile, critical, and recovery phase. The febrile phase lasts 2-8 days with symptoms like headache, nausea, rash, vomiting, myalgias, and arthralgias, and patients usually recover after this phase. The critical phase lasts 1-3 days and is manifested by thrombocytopenia, capillary permeability, and organ damage. The recovery phase lasts 3-6 days, and various complications develop, such as pulmonary edema and excessive intravenous fluid revive [66]. Dengue shock syndrome can also appear, which can get better with supportive treatment [67]. Dengue disease presents a wide range of clinical presentations, making diagnosis and therapy difficult. There is no distinction between events that will and will not develop into severe dengue, which is decided by the instant start of capillary aperture, either with or without hemorrhage, followed by thrombocytopenia, anemia, leucopenia, liver damage, and, in rare circumstances, death [54].

An evaluation of incubation time was made for the patients who had a blood transfusion and tested positive for Zika by RT-PCR, and clinical signs appeared 3-14 days after transfusion. Its symptoms include fever, conjunctivitis, vomiting, rash, arthritis, and headache, usually appearing after viremia. It is best to avoid using aspirin, NSAIDs, and steroids since the symptoms are clinically similar to those of the febrile phase of dengue [66]. The detection of Zika RNA in serum is restricted to the first five days of the disease, and after that, the serum accumulates [54].

**Fig. 3. Aedes species mosquito bites on the human body:** Aedes aegypti is the primary vector of chikungunya virus, zika virus, dengue virus, yellow fever virus and is endemic in various geographical areas. Various symptoms also appear after the incubation period of 7 days, like high fever, headache, rashes, nausea, pain behind the eyes, vomiting, and the mosquito saliva contains allergens, resulting in allergic reactions.

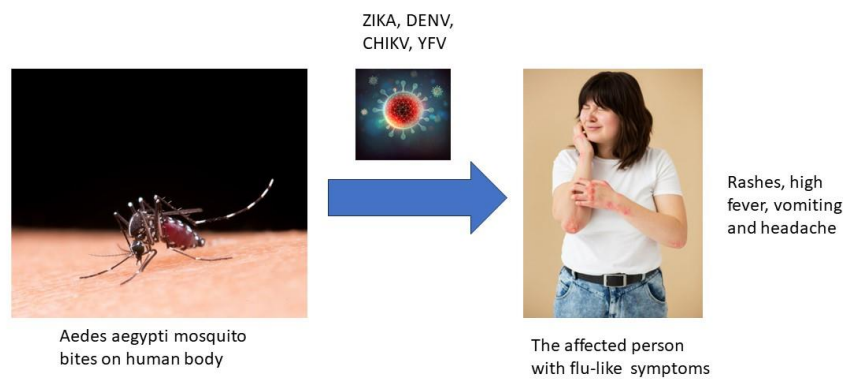

## 7. Diagnosis:

Laboratory studies should diagnose arbovirus infection due to various symptoms and colliding geographical distributions. Various serological diagnosis test includes indirect IgG assay, enzyme-linked immunosorbent assays (ELISA), IgM capture, magnetic immunoassay (MIA), rapid diagnostic test (RDT), immunofluorescence assay (IFA) [68, 69]. The serum neutralization test is most commonly used for serological diagnosis. ELISA detects IgM and IgG antibodies, particularly for arbovirus. IgM diagnoses recent infections, and IgG diagnoses older exposure by differentiating it from recent infections [68]. Serological assays are mainly used for diagnosis because it is rapid and inexpensive and fitter for diagnosing past and deadly infections, but various problems also appear like lack of specificity due to cross-reactivity between similar viruses, limited access to laboratory tools, the utilization of live viruses for antigen synthesis and plaque neutralization testing, false- negative result due to 3-5 days gap after initial infection before antibodies may be identified, and the price of doing separate experiments for each virus [70, 71, 69]

RT-PCR in serum is the primary test for identifying the viral Zika, chikungunya, and dengue virus during early stages. Diagnosis tests include nucleic acid sequence-based amplification (NASBA) and loop-mediated amplification (LAMP). For the diagnosis of various viral infections, such as respiratory, gastrointestinal, sexually transmitted, meningitis, and tropical diseases, RT-PCR has been recognized as the gold standard [69]. RT-PCR also detects nucleic acid in human and animal samples [72]. RT-PCR-specific serotypes can modify DENV serotypes even at severe stage [71]. Next-generation sequencing method has been utilized to detect arbovirus in patients with undetected meningitis/ encephalitis and in patients whose RT-PCR results are unclear. The molecular technique is mainly used for patients with acute phases of severe disease or etiological analysis of a disease outbreak of an unknown source. Recent price reductions for reagents like primers have allowed certain institutions with adequate funding to apply PCR-based methods for routine diagnosis [69].

The LAMP method, first presented in 2000, is used to amplify RNA or DNA at constant temperature using Bacillus stearothermophilus DNA polymerase 1. LAMP offers more benefits as compared to RT-PCR as its capacity to conduct a reaction at a single-set temperature, a reaction time of less than 15 minutes for the Ebola virus, and color change can be detected by unaided eyes to identify a positive reaction [69]. The serological test is usually conducted to quickly detect arbovirus diseases, supporting the commencement of control measures such as vaccination, vector control, and public awareness activities [72]. Immediately detecting arboviral infection is necessary for reporting virus activity in a specific area to ensure proper patient care. Because many arboviral diseases lack specialized vaccines or antiviral medications, early diagnosis is essential for enabling preventative steps during disease outbreaks, such as vector control methods, public awareness programs, and direct patient treatment courses.

8. **Vaccinations:**

According to the WHO, vaccination is the most effective preventive measure for controlling infectious diseases. In some countries like Asia and Africa, people use traditional medicines instead of vaccines for some diseases. CHIKV E3 and E2 proteins and Ankara virus-based vaccines have been used to protect different types of mice against CHIKV infection. DENV vaccines, like live attenuated (Dengvaxia), have been developed to protect all four serotypes.

DENV vaccines have been used in endemic regions. However, they cause severe diseases in people who never had dengue but were vaccinated, so currently, no other DENV vaccines are licensed for widespread use [73, 74]. Vaccination is not recommended for travelers but is suggested for children with previous DENV infection [48]. Curcumin, a common food ingredient, was used to reduce the spreading of ZIKV by suppressing the virus from attaching to cells without negatively affecting the viability of the cells [73].

Various vaccines like mouse brain-derived inactivated, cell culture-derived live-attenuated, and cell culture-derived inactivated vaccines are used to control pig and mosquito infection. In 1990, the Primary hamster kidney (PHK) that developed the inactivated vaccine was also used in China [75]. WNV vaccines have been developed and approved for horses, but no vaccine has been approved for humans [76]. Therapies used for the treatment of West Nile Virus (WNV) are polyclonal immune globulin (IVIGs), interferon, ribavirin, recombinant humanized monoclonal antibody, recipient of solid organ transplant, and corticosteroids [77]. The first licensed live-attenuated vaccine, YFV-17D, was introduced in the 1930s to control YFV, providing long-term immunity after a single dose and infecting Aedes—Aegypti midgut [76, 78]. A lipid nanoparticle (LNP) is a vaccine made of organic polymers, 100nm in size, that is used for treating inherited conditions [79]. Insect-specific virus (ISV) vaccines are developed to protect against mosquitoes and to produce chimeric vaccines. Licensed vaccines are not widely available for many arbovirus diseases, and research is ongoing to develop novel vaccines [80].

## 9. Prevention and control strategies:

Arbovirus infection has no specific treatment plan, but various preventive measures should be taken to reduce its symptoms, as shown in **Fig 4**. Nanotechnology can be used to improve the drug delivery and pharmaceutical qualities of therapeutic medicines, such as improved pharmacokinetic profiles and increased solubility. These drug delivery systems (DDS) include nano-emulsions (NE), which are a double-phase diffusion of two liquids, such as oils and water, sustained by an internally assembled surface-active agent [81]. Pain relievers like Paracetamol, ibuprofen, and antipyretics treat high fever and anti-inflammatory drugs for joint pain [73]. Arbovirus diseases can be controlled by early detection of viruses and monitoring through various surveillance activities. The surveillance program can detect new species development and increased transmission threats [72]. Various preventive measures are also taken to control arboviral diseases, such as the use of insecticide spray to kill mosquitoes and control the pyrethroid-resistance vector and the use of chemical and microbial larvicides and bacterial toxins to control larval growth [42].

**Fig 4: Prevention strategies for the control of various arboviruses**

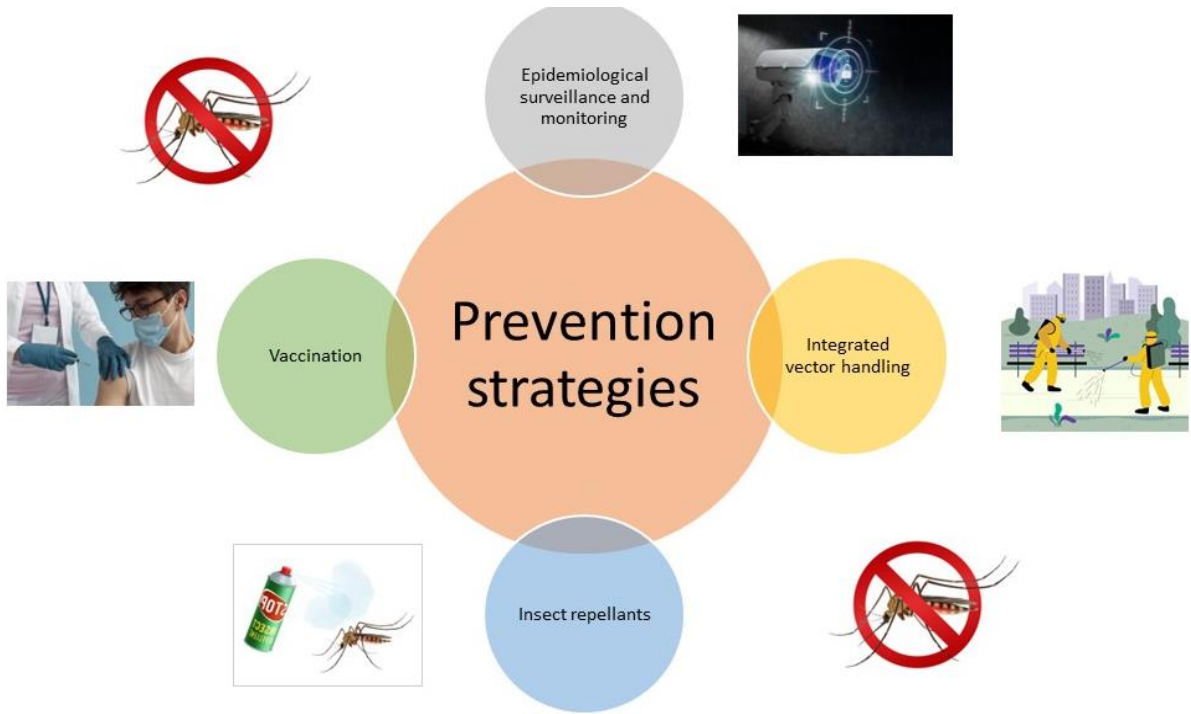

VeCtor population is also controlled by placing inter-dependent bacteria in vectors that un-function their reproductive cycle and restrict the maturing of vector species [82]. Wolbachia in most mosquitos reduces the mosquito population and helps in biological control [52]. Spatial repellents control human-vector contact and reduce malarial infection in humans. Traps eliminate vectors with a lethal combination of heat, moisture, carbon dioxide, and light, which attract the host. Insecticide-treated material (ITM) in clothes is used to protect from mosquito bites. This preventive measure reduces the incidence of vector-borne diseases like malaria and leishmaniasis [83]. Urban areas that COVID-19 will impact should lower their risk of yellow fever by removing possible mosquito breeding grounds and spraying larvicides to water storage containers and other locations where standing water gathers [10].

## 10. Conclusion:

The most discussed arboviral diseases in this article are DENV, ZIKV, and CHIKV. These arboviruses are producing epidemics in various countries, resulting in morbidity and complications, which increases the need for medical care and support services [54]. The RNAi pathway is the antiviral vector used to control arbovirus infection [84]. Enhancing our understanding of virus-vector interactions, which are more developed for mosquitoes than for ticks, will enable us to develop more accurate forecasts of arbovirus emergence and to put appropriate vector control strategies into place. The ongoing Dengue and Chikungunya virus outbreak and the risk of Zika and yellow fever virus should alert the governments, foundations, and WHO to speed up the research for mosquito-transmitted diseases [57]. Various arboviral diseases arise during different outbreaks, making managing these viruses more challenging. After decades of study, vaccines are unavailable for the most critical arboviral disease-causing dengue virus [73]. Arbovirus spread by Aedes mosquito poses a severe threat to public health and will need various preventive measures to be effectively maintained. Therefore, developing and reviewing alternate vector-control products and approaches is essential due to the growing spectrum of arboviruses [83]. Many arbovirus diseases lack specialized vaccines and antiviral medications, so early diagnosis is essential for providing preventive steps during disease outbreaks. The prevention strategies for vector populations include biological control, traps, insect repellants, monitoring programs, and ITM [52, 43]. Comprehensive knowledge about the arbovirus vaccines should be deepened to notify research directions, secure funding, and create a disease control policy for disease prevention and control [85]. This article provides a brief overview of the impact of various arboviral diseases and their epidemiology, transmission cycle, signs and symptoms, diagnosis, treatment, and preventive/ control measures.